\documentclass[aps,preprint]{revtex4}

\usepackage{amsmath,amssymb,amsfonts,graphicx,hyperref,amsthm,xcolor}
\newcommand{\beq}{\begin{equation}}
\newcommand{\eeq}{\end{equation}}
\newcommand{\bea}{\begin{eqnarray}}
\newcommand{\eea}{\end{eqnarray}}

\begin{document}

\title{Variations in the fine structure constant constraining gravity theories}

\author{V. B. Bezerra}
\email{valdir@fisica.ufc.br}\address{Departamento de F\'{i}sica, Universidade Federal da Para\'{i}ba, Caixa Postal 5008, CEP 58051-970, Jo\~{a}o Pessoa, PB, Brazil}

\author{M. S. Cunha }
\email{marcony.cunha@uece.br}\address{Grupo de F\'isica Te\'orica (GFT), Centro de Ci\^encias e Tecnologia, Universidade Estadual do Cear\'a, CEP 60740-000, Fortaleza, Cear\'a, Brazil.}

\author{C. R. Muniz}
\email{celio.muniz@uece.br}\address{Grupo de F\'isica Te\'orica (GFT), Universidade Estadual do Cear\'a, Faculdade de Educa\c c\~ao, Ci\^encias e Letras de Iguatu, Iguatu, Cear\'a, Brazil.}

\author{M. O. Tahim}
\email{makarius.tahim@uece.br}\address{Grupo de F\'isica Te\'orica (GFT), Universidade Estadual do Cear\'a, Faculdade de Educa\c c\~ao, Ci\^encias e Letras do Sert\~ao Central, Quixad\'a, Cear\'a, Brazil.}

\author{H. S. Vieira}
\email{horacio.santana.vieira@hotmail.com}
\address{Departamento de F\'{i}sica, Universidade Federal da Para\'{i}ba, Caixa Postal 5008, CEP 58051-970, Jo\~{a}o Pessoa, PB, Brazil and\\Centro de Ci\^{e}ncias, Tecnologia e Sa\'{u}de, Universidade Estadual da Para\'{i}ba, CEP 58233-000, Araruna, PB, Brazil}

\begin{abstract}

In this paper, we investigate how the fine structure constant, $\alpha$, locally varies in the presence of a static and spherically symmetric gravitational source. The procedure consists in calculating the solution and the energy eigenvalues of a massive scalar field around that source, considering the weak-field regimen. From this result, we obtain expressions for an spatially variable fine structure constant by considering suitable modifications in the involved parameters admitting some scenarios of semi-classical and quantum gravities. Constraints on free parameters of the approached theories are calculated from astrophysical observations of the emission spectra of a white dwarf. Such constraints are finally compared with those ones obtained in the literature.\\

\vspace{0.75cm}
\end{abstract}

\maketitle

\section{Introduction}

How \textit{constant} are the constants of nature? Which ones are really constant? These are some issues formulated by various researchers since Planck proposed in the early twentieth century a natural system of units for measurements of space, time, mass, and their derivations. Such units do not rely on arbitrary standards, rather they are built from the universal realms of electromagnetism, gravity and quantum mechanics. This one as well as other fundamental system of units are discussed in \cite{Wilck}.

In this context, the constants of nature which are dimensionless acquire great relevance. One of the most important of them is the fine structure constant, $\alpha$, given by the combination of the elementary charge $e$, reduced Planck constant $\hbar$, and light velocity in vacuum $c$, $\alpha=e^2/\hbar c\approx 1/137$. It is currently measured with high precision, indicating the strength of the electromagnetic interactions, and was introduced for the first time by A. Sommerfeld \cite{Sommer} in order to explain the fine structure of the hydrogen atom spectra. Since then one seeks knowing if such a constant has varied in the time and space, principally from astrophysical \cite{Barrow,Bambi} or cosmological observations \cite {Bahcall,Barrow1} (and references therein).

From a theoretical point of view, some researches have investigated wider consequences of a variable fine-structure constant, including those ones based on cosmic strings \cite{Nasseri1,Nasseri2}, grand unification theories \cite{Marciano,Langacker}, and on extra dimensions \cite{Youm,Palma}. Other theories \cite{Barrow2} also predict that the fine-structure constant became practically constant since the universe entered in its prevailing accelerated expansion epoch, meaning that its temporal variation has been negligible since then. On the other hand, spatially variable $\alpha$ implies that the laws of physics can be different at different locations of the Universe. The experimental support for global ({\it i.e.}, cosmological) variations of $\alpha$ must be confirmed by other Earth-based tests \cite{Berengut}.

This paper deals with local variations of the fine structure constant induced by a spherically symmetric gravitational source, taking into account different scenarios, as for example the one which includes the presence of a self-force, and different approaches which consider Generalized Uncertainty Principle (GUP), non-commutative space, asymptotically safe gravity, and Horava-Lifshitz gravity. The Schwarzschild solution has been assumed here as a background metric, although static and spherically symmetric solutions in alternative theories of gravity must be described by other geometries. In these alternative theories the modifications in general relativity are Planck scale suppressed in low energies. Thus, our procedure considers that the Schwarzschild solution is slightly corrected in its proper parameters, such as Newtonian gravity constant or the source mass. These parameters explicitly appear in the eigenvalues of the energy found for a massive scalar field in the gravitational field of the source that we are considering. From their modification we obtain an effective Planck constant and as a consequence a spatially variable $\alpha$. Then, we will calculate from astrophysical measurements of $\alpha$, specifically from data of the emission spectra of white dwarfs found in the literature \cite{Barrow}, the constraints on the parameters arising in the scenarios which we will consider.  There are a wide variety of theories which seek quantize the gravity, so that only an experimental investigation can select those ones more consistent, while a true quantum theory of gravity awaits its discovery. The present paper can contribute to this investigation, by proposing that certain astrophysical measurements of $\alpha$ fix the values permitted for some parameters of those theories.

The paper is organized as follows: In section II we obtain the solution of a massive scalar field in the Schwarzschild metric, and determine the energy spectrum. In section III, we calculate the effective fine structure constant in the aforementioned scenarios. Finally, in section IV, we draw the conclusions and close the paper.

\section{Scalar Field Around a Static Spherically Symmetric Source}

The scalar massless field in the background of the Schwarzschild solution was discussed during last decades \cite{Frolov} (See references therein about the earlier works on this subject). It is important to call attention to the fact that all these discussions considered only approximate solutions to the problem. In fact, only recently the exact and complete solutions for the field dynamics of a scalar field in Schwarzschild spacetime was solved in \cite{Horacio1,Horacio2} in terms of the Heun's functions due to advances on the study these functions in recent years.

We will now briefly summarize the investigations concerning the behavior of massive scalar fields in the gravitational field of a static black hole, considering solutions in the region exterior to the horizon. To do this, we must solve the covariant Klein-Gordon equation, which is the equation that describes the behavior of scalar fields in the spacetime under consideration. In a curved spacetime, we can write the Klein-Gordon equation of an uncharged massive scalar particle coupled minimally with gravity as
\begin{equation}
\left[\frac{1}{\sqrt{-g}}\partial_{\mu}\left(g^{\mu\nu}\sqrt{-g}\partial_{\nu}\right)+m^{2}\right]\Psi=0\ ,
\label{eq:Klein-Gordon_cova}
\end{equation}
where we adopted the natural units $c \equiv \hbar \equiv 1$ and $m$ is the particle mass. On the other hand, the background generated by a static and uncharged black hole is represented by the Schwarzschild metric \cite{MTW:1973}, which in the Boyer-Lindquist coordinates \cite{JMathPhys.8.265} can be written as
\begin{equation}
ds^{2}=f(r)dt^{2}-f(r)^{-1}dr^{2}-r^{2}d\Omega^2,
\label{eq:metrica_Kerr-Newman}
\end{equation}
where $f(r)=\left(1-\frac{2M}{r}\right)$, $d\Omega^2=d\theta^{2}+\sin^{2}\theta\ d\phi^{2}$ and $M$ is the mass of the source, with $G \equiv 1$. In order to solve Eq.~(\ref{eq:Klein-Gordon_cova}) in the background given by Eq.~(\ref{eq:metrica_Kerr-Newman}), we assume that its solution can be separated as follows
\begin{equation}
\Psi=\Psi(\mathbf{r},t)=R(r)Y_{l}^{m_l}(\theta,\phi)\mbox{e}^{-i\omega t}\ ,
\label{eq:separacao_variaveis}
\end{equation}
where $Y_{l}^{m_l}(\theta,\phi)$ are the spherical harmonic functions. Plugging Eq. (\ref{eq:separacao_variaveis}) and the metric given in Eq.~(\ref{eq:metrica_Kerr-Newman}) into (\ref{eq:Klein-Gordon_cova}), we obtain the following radial equation
\begin{equation}
\frac{d}{dr}\left[r(r-2M)\frac{dR}{dr}\right]+\left(\frac{r^{3}\omega^{2}}{r-2M}-m^{2}r^{2}-\lambda_{lm_l}\right)R=0\ ,
\label{eq:mov_radial_1}
\end{equation}
where $\lambda_{lm_l}=l(l+1)$.

This equation has singularities at $r=(a_{1},a_{2})=(0,2M)$, and at $r=\infty$, and can be transformed into a Heun-type equation by using the coordinate transformation in the independent variable 
\begin{equation}
x=\frac{r-2M}{2M},
\label{eq:homog_subs_radial}
\end{equation}
and in the dependent one by introducing a new function, $Z(x)$, such that $R(x)=Z(x)[x(x-1)]^{-1/2}$. Then, we can transform Eq.~(\ref{eq:mov_radial_1}) into an equation for $Z(x)$, which is written as
\begin{equation}
\frac{d^{2}Z}{dx^{2}}+\left[A_{1}+\frac{A_{2}}{x}+\frac{A_{3}}{x-1}+\frac{A_{4}}{x^{2}}+\frac{A_{5}}{(x-1)^{2}}\right]Z=0 ,
\label{eq:mov_radial_x_heun}
\end{equation}
with the coefficients $A_{1}$, $A_{2}$, $A_{3}$, $A_{4}$, and $A_{5}$ given by
\bea
A_{1}&=&4 M^2 \left(\omega ^2-m^2\right); \label{eq:A1_mov_radial_x_normal}\\
A_{2}&=&\frac{1}{2}+\lambda_{lm_l} +4 M^2 \left(m^2-2 \omega ^2\right)\ ;
\label{eq:A2_mov_radial_x_normal}\\
A_{3}&=&-\frac{1}{2}-\lambda_{lm_l};
\label{eq:A3_mov_radial_x_normal}\\
A_{4}&=&\frac{1}{4}+4 M^2 \omega ^2;
\label{eq:A4_mov_radial_x_normal}\\
A_{5}&=&\frac{1}{4}.
\label{eq:A5_mov_radial_x_normal}
\eea
The Eq.~(\ref{eq:mov_radial_x_heun}) is a confluent Heun equation in the normal form \cite{Horacio1} whose solutions are the confluent Heun funtions. Therefore, the general L. I. solutions over the entire range $0\leq x < \infty$ can be written as 
\bea
R(x) = C_{1} \mbox{e}^{\frac{1}{2}\alpha x}x^{\frac{1}{2}\beta} \mbox{HeunC}(\alpha,\beta,\gamma,\delta,\eta;x)+
C_{2}\mbox{e}^{\frac{1}{2}\alpha x}\ x^{-\frac{1}{2}\beta}\ \mbox{HeunC}(\alpha,-\beta,\gamma,\delta,\eta;x),~~~~
\eea \label{eq:solucao_geral_radial_Kerr-Newman_gauge}%
where $C_{1}$ and $C_{2}$ are constants, and the parameters $\alpha$, $\beta$, $\gamma$, $\delta$, and $\eta$ are given from of the coefficients of Eq.~(\ref{eq:mov_radial_x_heun}) above by
\begin{subequations}
\bea
\alpha &=& - 4M\left(m^{2}-\omega^{2}\right)^{1/2};
\label{eq:alpha_radial_HeunC_Kerr-Newman}\\
\beta&=&i4M\omega;
\label{eq:beta_radial_HeunC_Kerr-Newman}\\
\gamma&=&0;
\label{eq:gamma_radial_HeunC_Kerr-Newman}\\
\delta&=&4M^{2}\left(m^{2}-2\omega^{2}\right);
\label{eq:delta_radial_HeunC_Kerr-Newman}\\
\eta&=&-l(l+1)-4M^{2}\left(m^{2}-2\omega^{2}\right).
\label{eq:eta_radial_HeunC_Kerr-Newman}
\eea
\end{subequations}

These two functions are linearly independent solutions of the confluent Heun differential equation provided that $\beta$ is not an integer \cite{Ronveaux}.

We must now calculate the field energy eigenvalues, imposing boundary conditions on the solutions at the asymptotic region (infinity), which in this case, requires the necessary condition for $R(x)$ to be a polynomial, since the confluent Heun solutions have irregular singularities there. The confluent solution is defined in the disk $|z|<1$ by the series expansion
\bea
\text{HeunC}(\alpha,\beta,\gamma,\delta,\eta,z)=\sum_{n=0}^\infty v_{n}(\alpha,\beta,\gamma,\delta,\eta)z^n,
\label{HeunC}
\eea
together with the condition $\text{HeunC}(\alpha,\beta,\gamma,\delta,\eta,0)=1$.
The coefficients $v_{n}(\alpha,\beta,\gamma,\delta,\eta)$ are determined by three-term recurrence relation ($n>0$)
\beq
A_{n}v_{n}=B_{n}v_{n-1}+C_{n}v_{n-2}
\label{recurrence}
\eeq
with initial condition $v_{-1}=0,\,\,v_{0}=1$ \cite{Fisiev}. Here
\bea
\hskip -.truecm A_{n}&=&1+{\frac{\beta}{n}} \\
\hskip -.truecm B_{n}&=&1+{\frac{-\alpha+\beta+\gamma-1}{n}}+{\frac{\eta-(-\alpha+\beta+\gamma)/2-\alpha\beta/2+\beta\gamma/2}{n^2}} \\
\hskip -.truecm C_{n}&=&{\frac{\alpha}{n^2}}\left({\frac \delta \alpha}+{\frac {\beta+\gamma}{2}}+n-1\right).
\label{rec_coeff}
\eea
Thus, in order to have a polynomial solution of the confluent Heun function we must impose the so called $\delta_N$ and $\Delta_{N+1}$ conditions, namely  \cite{Ronveaux},
\begin{eqnarray}
\frac{\delta}{\alpha}+\frac{\beta+\gamma}{2}+1&=&-N\label{eq:cond_polin_1}\\
\Delta_{N+1}&=&0
\label{eq:cond_polin_2}
\end{eqnarray}
with $N$ being a non-negative integer. The first condition Eq. (\ref{eq:cond_polin_1}) gives the following expression for the energy levels 
\begin{equation}
N+1+\frac{i4M\omega}{2}-\frac{4M^2\left(m^2-2\omega^{2}\right)}{4M\sqrt{m^{2}-\omega^{2}}}=0.
\label{eq:energy_levels}
\end{equation}

Now we consider in our analysis the low frequency limit, $\omega M\rightarrow 0$, which means that the particle is not absorbed by the black hole. In fact, the relative absorption probability of the scalar wave at the event horizon surface of a static and uncharged black hole is given by \cite{Horacio1}
\begin{equation}
\Gamma_{ab}=1-e^{-8\pi M\omega},
\end{equation}
and, in the considered limit, this probability goes to zero. In this regime, the particle does not penetrate via tunneling the effective potential barrier that exists around the black hole (the Regge-Wheeler potential), in such a way that the reflection coefficient tends to one \cite{Frolov}. The considered limit also can be thought as a cutoff introduced in order to eliminate high frequencies. It is worth yet notice that, for $m=0$ and in that approximation, we have the solution $n=-1$, which does not make sense. This seems to suggest that, while classically stable orbits for such a kind of particle exist just at $r=3M$ \cite{Frolov}, quantum mechanically they do not exist in any way, {\it i.e.}, there is no massless stationary states completely exterior to the black hole.

We take now into account the possible stationary states which are totally exterior to the black hole, by neglecting the terms $\mathcal{O}(M\omega)$ in Eq. (\ref{eq:energy_levels}) according to the aforementioned low frequency limit. Then we have
\begin{subequations}
\beq
N+1 - \frac{Mm^2}{\sqrt{m^2-\omega^2}}\approx0,
\eeq%
which implies
\beq
(N+1)^2\approx\frac{M^2m^4}{m^2-\omega^2},
\eeq
\end{subequations}
and thus we easily get the energy eigenvalues, given by
\beq
\label{03}
E_{n}\approx m c^2\sqrt{1-\frac{G^{2}m^{2}M^{2}}{\hbar^2c^2n^2}},
\eeq
where we have used $E_n=\hbar\omega_n$, and reintroduced $G$, $\hbar$, and $c$. Furthermore in the above expression $n=N+1$ is redefined such that the states begin from $n=1$. The Eq. (\ref{03}) was also found in \cite{Barranco} by following a different approach.

Notice that the bound energy $E_{n,b}=E_n-mc^2$ reaches a minimum equal to $-mc^2$ when $GM/c^2=n\hbar/mc$ or $r_S/2=n\lambda_C$, where $r_S$ is the Schwarzschild radius and $\lambda_C$ is the Compton wave length of the particle. Thus, in order to consider all the quantum states of the particle, {\it i.e.}, without a bounded $n$, we must have $\lambda_C\geq r_S/2$, which is compatible with the low frequency limit that we are using, implying real energies and therefore without the possibility of tunnel effect to the black hole interior. Such exterior solutions are quite suitable to our purposes and will be applied to spherically symmetric sources which are not black holes.

We have reintroduced the fundamental constants in Eq. (\ref{03}) in order to call attention to the fact that its non-relativistic approximation, $\mathcal{O}(1/c^2)\rightarrow 0$, after subtracting the rest energy of the particle, is exactly equal to the Bohr's energy levels of a ``gravitational hydrogen-like atom'', namely
\begin{eqnarray}\label{Bohrlevels}
\label{03.1}
E_{n}\approx-\frac{G^2m^3M^2}{2\hbar^2n^2},
\end{eqnarray}
which also was obtained in \cite{Laptev,Doran}. This result will be henceforth used for our purposes of calculating the variable fine structure constant according to some approaches to be considered below.

\section{On local variations of the fine structure constant}

In this section we will investigate the spatial variations of the fine structure constant according to semi-classical and quantum theories of gravity, obtaining constraints on the involved parameters from the current astrophysical measurements, specifically from emission spectra of white dwarfs.
\subsection{Scenario I: Existence of a self-force on the particle}

According to \cite{Burko}, for  any  circular orbit around a static spherical gravitational source and in the approximations considered so far, a scalar particle with charge $e$ suffers a repulsive radial force which is given by
\begin{equation}\label{self}
F^{self}\approx ke^2(G^3/c^6)M^2\Omega^2/r^2=k e^2(G^3/c^6)M^2L^2/m^2r^6,
\end{equation}
where $\Omega$ is the angular velocity of the particle, $L=mvr$ is its angular momentum, and $k$ a undetermined numerical dimensionless constant. Taking into account the Newtonian gravity and the second law of classical mechanics, we have
\begin{equation}\label{secondlaw}
\frac{L^2}{mr^3}=\frac{GMm}{r^2}-k \frac{e^2G^3M^2L^2}{m^2c^6r^6},
\end{equation}
from which
\begin{equation}\label{Bohrradii}
r=\frac{n^2\hbar^2}{GMm^2}\left(1+k\frac{e^2G^3M^2}{mc^6r^3}\right),
\end{equation}
where the quantization rule $L=n\hbar$ was taken into account. Notice that the expression (\ref{Bohrradii}) is compatible with (\ref{Bohrlevels}) when $k=0$, {\it i.e.}, for the gravitational Bohr orbits. From Eq. (\ref{Bohrradii}), we can define an effective Planck's constant, $\hbar_{eff}=\hbar(1+k e^2G^3M^2/mc^6r^3)^{1/2}$, and thus find a new expression for the fine structure constant $\alpha=e^2/\hbar c$, namely, an effective fine structure constant, $\alpha_{eff}$, given by
\begin{equation}\label{alpha}
\alpha_{eff}=\alpha\left(1+k \frac{e^2G^3M^2}{mc^6r^3}\right)^{-1/2}\approx \alpha\left(1-k \frac{e^2G^3M^2}{2mc^6r^3}\right),
\end{equation}
in which we notice that $\alpha_{eff}\leq \alpha$.

Constraints on the value of $k$ can be found from measurements of $|\Delta\alpha/\alpha|=|[\alpha(r)-\alpha]/\alpha|\thicksim 10^{-5}$ made using the spectra of white dwarfs \cite{Barrow}. Thus, we have that
\begin{equation}\label{constraint alpha}
\frac{|\Delta\alpha|}{\alpha}=\frac{|\alpha_{eff}-\alpha|}{\alpha}\gtrsim k\frac{e^2G^3M^2}{2mc^6r^3},
\end{equation}
or
\begin{equation}
k\lesssim 4.4\times 10^{48},
\end{equation}
where it was used $r=0.022R_{\odot}$ and $M=0.51M_{\odot}$ \cite{Barrow}. A similar analysis on the variation of the fine structure constant considering the self force on a particle in the spacetime of a cosmic string was made in \cite{Nasseri1}.

\subsection{Scenario II: Generalized Uncertainty Principle (GUP)}

String inspired theories and black hole physics predict the existence of a fundamental length in the Universe. A generalization of the Heisenberg principle, namely, GUP, incorporates this fact \cite{Xiang}. Such a principle is very closely linked to the Doubly Special Relativity (DSR), where modified dispersion relations (or MDR) prevail in order to yield an additional invariant scale (besides that one related to the speed of light), which matches the scale at which the quantum nature of the spacetime manifests itself \cite{Hinter, Mairi}. In other words, GUP and DSR (with MDR) are different aspects of the same idea, envisaging the existence of an observer-independent minimal length scale or a maximum momentum in quantum gravity theory.

Specifically, GUP is often written as
\begin{equation}\label{GUP}
\Delta x_i\Delta p_i\geq \hbar\left[1+\beta^2L_P^2\left(\frac{\Delta p_i}{\hbar}\right)^2\right],
\end{equation}
where $L_p=\sqrt{\hbar G/c^3}$ is the Planck length and $\beta$ a dimensionless constant. The minimal uncertainty for $x_i$ is $\Delta x_i^{(min)}=2\beta L_P$, considering a free particle. For the particle around a spherically symmetric gravitational source, we suppose that this minimum corresponds to $a_{GB}(r)$, which is the gravitational Bohr's radius obtained when $n=1$ into Eq. (\ref{Bohrradii}). Thus, we can write the previous expression for $\Delta p_i$ in terms of this radius, which results in an effective $\hbar_{eff}$ given by
\begin{equation}\label{Delta p_i}
\hbar_{eff}=\hbar\left[1+\frac{a_{GB}^2}{4\beta^2L_P^2}\left(1-\sqrt{1-\frac{4\beta^2L_P^2}{a^2_{GB}}}\right)^2\right].
\end{equation}
If we consider that $L_P/a_{GB}\ll 1$, we have the approximation $\hbar_{eff}\approx(1+ \beta^2L_P^2/a^2_{GB})$, which combined with the explicit expression for $a_{GB}$ results in an effective fine structure constant given by
\begin{equation}\label{GUP alpha}
\alpha_{eff}\approx  \alpha\left[1-\beta^2\left(\frac{G^2Mm^2}{c^3\hbar}-k\frac{e^2G^5M^3m}{c^9\hbar r^3}\right)\right].
\end{equation}
Neglecting the second term in parenthesis of Eq. (\ref{GUP alpha}), we can infer a constraint on $\beta$ using the same data of the previous scenario. In doing this procedure, we get the following constraint
\begin{equation}\label{beta}
\beta\lesssim 2.3 \times 10^{17}.
\end{equation}

\subsection{Scenario III: Non-commutative geometry}

In this case, we take into account that point-like masses are diffused throughout space due to a fundamental uncertainty with respect to its localization. In other words, the coordinates of a particle obey a non-commutative relation. Thus, its mass density is distributed according to a Gaussian function instead of a Dirac one, in the following form \cite{Alavi}
\begin{equation}\label{Gaussian density}
\rho(r)=\frac{M}{(4\pi\theta)^{3/2}}\exp{\left(-\frac{r^2}{4\theta}\right)},
\end{equation}
 where $\theta$ is the non-commutativity parameter. Integrating this expression over the volume up to $r$, we find that the point-like source mass is
 \begin{equation}\label{non-commutative mass}
M(r)=\frac{2M}{\sqrt{\pi}}\gamma\left(\frac{3}{2},\frac{r^2}{4\theta}\right),
\end{equation}
 where $\gamma(3/2,x)=\int_0^xt^{1/2}e^{-t}dt$ is the lower incomplete Gamma function. It is worth point out that $M(r)\rightarrow M$ when $\theta\rightarrow 0$.

Now, we consider that the gravitational source has a mass given by Eq. (\ref{non-commutative mass}). Then inserting it into Eq. (\ref{Bohrlevels}), we can define an effective Planck constant, $\hbar_{eff}$, such that $\hbar_{eff}=\hbar \frac{\sqrt{\pi}}{2}\left[\gamma\left(\frac{3}{2},\frac{r^2}{4\theta}\right)\right]^{-1}$. Thus, using this result, we find a new fine structure constant, $\alpha_{eff}$, written as
\begin{equation} \label{non-commutative alpha}
\alpha_{eff}=\alpha\left[\frac{2}{\sqrt{\pi}}\gamma\left(\frac{3}{2},\frac{r^2}{4\theta}\right)\right].
\end{equation}
Doing the appropriate expansion of $\gamma(3/2,r^2/4\theta)=\sqrt{\pi}/2+e^{-r^2/4\theta}[-r/2\sqrt{\theta}+\mathcal{O}(\sqrt{\theta}/r)]$, we have the following approximate result
\begin{equation}\label{non-commutative alpha 2}
 \alpha_{eff}\approx \alpha\left(1-e^{-r^2/4\theta}\frac{r}{\sqrt{\pi\theta}}\right),
 \end{equation}
for $\sqrt{\theta}\ll r$. In order to establish constraints on the $\theta$ parameter by taking into account the astrophysical measurements about $|(\Delta\alpha)|/\alpha\sim 10^{-5}$ previously considered, we can graphically obtain
 \begin{equation}\label{constraint on theta}
 \sqrt{\theta}\lesssim 0.24 r,
 \end{equation}
which is about $25\%$ of the source radius. Considering $r=0.022R_{\odot}$, as before, we have that $\sqrt{\theta}\lesssim 3.7\times 10^6$ m.

 \subsection{Scenario IV: Asymptotic safety gravity}

The procedure adopted in this and the following section is based on an extension of models in which the relative spatial variation of $\alpha$ by considering two distinct points is proportional to the difference between the Newtonian gravitational potentials in these places \cite{Lip} (e.g., white dwarf surface and Earth), {\it i.e.}, $\Delta\alpha/\alpha=k_\alpha\Delta\phi_N$, where $\phi_N=GM/c^2r$. The generalization proposed here is to consider $\Delta\alpha/\alpha=k_\alpha\Delta\Phi$ where $\Phi$ is the post-Newtonian potential so that $e^{2\Phi}=g_{00}$ \cite{Rindler}. Thus we have
\begin{equation}\label{correctalpha}
\frac{\Delta\alpha}{\alpha}=k_\alpha\Delta\Phi\approx k_\alpha(\Delta\phi_N+\phi_{cS})>\phi_{cS},
\end{equation}
where $\phi_{cS}$ is the term that corrects the $g_{00}$ coefficient regarding the Schwarzschild metric in the theories to be considered.

For the exterior region of a static black holes it was proposed \cite{Bonanno} (see references therein) the existence of an effective improved solution based on the asymptotic safety approach, which incorporates quantum corrections to the classical solution (\ref{eq:metrica_Kerr-Newman}). That effective solution is equivalent to put in this latter a corrected Newtonian gravity constant given by
\begin{equation}\label{improved G}
G(r)=G\left[1-\tilde{\omega}\frac{L_P^2}{r^2}+\mathcal{O}\left(\frac{L_P^3}{r^3}\right)\right],
\end{equation}
where we have again $L_p=\sqrt{\hbar G/c^3}$ and $\tilde{\omega}$ is a dimensionless constant. The latter is calculated according to canonical perturbative quantization of Einstein's gravity, which precisely yields $\tilde{\omega}=167/30\pi$. However, as it is known, such a canonical quantization leads to a non-renormalizable theory. We prefer to obtain an experiment-based constraint on the new length $L\backsim\sqrt{\tilde{\omega}}L_P$ that defines the scale under which the theory is valid. Neglecting terms $\mathcal{O}(L^3/r^3)$ in Eq. (\ref{improved G}), the term that corrects the Schwarzschild metric is $\phi_{cS}=GML^2/r^3$, from which we obtain
\begin{equation}\label{nonlocal alpha}
\frac{\Delta\alpha}{\alpha}>\frac{GML^2}{c^2r^3}.
\end{equation}
The constraint on $L$ comes from the measurements on $|\Delta\alpha|/\alpha$ under consideration, which permits us to establish the following constraint
\begin{equation}\label{constraint omega}
L< 3.16 \times 10^{-2} \sqrt{\frac{c^2r^3}{GM}}.
\end{equation}
If we use the radius and mass used so far, this upper bound is $L<7.0\times 10^{7}$ m.

\subsection{Scenario V: Ho\v{r}ava-Lifshitz gravity}

The theoretical formulation known as Ho\v{r}ava-Lifshitz gravity is a recent attempt to build a renormalizable quantum theory of gravity \cite{Horava1, Horava2, Horava3}.  This formulation has a manifest broken Lorentz symmetry at ultraviolet scale which occurs via anisotropic scaling of space and time. Such a theory reduces to general relativity in the appropriated (infrared) limit. In this context, there exists a static and spherically symmetric general solution of Ho\v{r}ava-Lifshitz gravity, which in the infrared approximation is given by Eq. (\ref{eq:metrica_Kerr-Newman}), with $f(r)$ corrected to \cite{Celio}
\begin{equation} \label{H-L f(r)}
f(r)\approx \left(1-\frac{2GM}{c^2r}+\frac{2\hbar^2G^2M^2}{\omega c^6r^4}\right),
\end{equation}
where $\omega$ is a free parameter of the Ho\v{r}ava-Lifshitz theory. The limit $\omega\rightarrow \infty$ corresponds to the infrared regime, where this theory goes to general relativity.

As the aforementioned metric represents a slight correction to the Schwarzschild one, we can assume that such a correction leads to a relative spatial variation in the fine structure constant so that
\begin{equation}\label{H-L alpha}
\frac{\Delta\alpha}{\alpha}>\frac{\hbar^2 G^2 M^2}{\omega c^6 r^4}.
\end{equation}
Considering once more the data used in order to obtain constraints in the involved parameters of the theories examined until now, we have that $\omega > 10^{-134}$ m$^{-2}$.

\section{Closing remarks}

In summary, we have initially presented analytic solutions of the radial part of the Klein-Gordon equation for an uncharged massive scalar field in the spacetime of a static and spherically symmetric source (Schwarzschild's spacetime). Then, we have drawn the energy eigenvalues of that field in the regimen in which $M \omega \rightarrow 0$, so that the particle does not tunnel into the black hole, {\it i.e.}, we have considered only the completely exterior solutions. The consistence of these calculations was shown via non-relativistic limit in which we obtained the gravitational analogue of the Bohr's levels for the hydrogen atom. With this, we justify the use of quantized orbits in our purpose of finding the local variations of the fine structure constant.

Next, we investigated how the fine structure constant $\alpha$ depends on the parameters of some semi-classical and quantum theories of gravity. At this stage, we analyzed how modifications in the Newtonian constant $G$ (in the scenarios which consider particle self-force, asymptotically safety and Ho\v{r}ava-Lifshitz gravities), as well as in the source mass (through the non-commutative geometry) or directly in the proper Planck constant $\hbar$  (via GUP), from their insertion into Eq. (\ref{Bohrlevels}), imply a variation in this last constant, and as a consequence in the fine structure constant. This found effective fine structure constant became a function of the radial coordinate calculated from the source, and measurements of $\alpha_{eff}$ based on the emission spectra of white dwarfs allowed us inferring constraints on the undetermined parameters of those approaches. It is worth noticing that all the analysis made here showed up that $\alpha_{eff}$ is such that $\alpha_{eff}<\alpha$, which means the its values are lower than the value of the fine structure constant outside the scenarios considered.

The table I resumes the results, which are compared with the ones registered in the literature and collected from similar astrophysical measurements, when they exist. From this table, we can see that the best parameter constraints obtained were those ones related to GUP and non-commutative space scenarios, and the worst was the one calculated via Ho\v{r}ava-Lifshitz gravity. For the constraints obtained using the particle self-force and the asymptotic safety gravity approach, we have no results in the literature, as far as we know, in order to make comparisons.
\begin{table}[h]
\caption{Parameter constraints}
\centering
\begin{tabular}{|l|c|c|c|}\hline
{\bf Scenario} & {\bf Parameter} & {\bf Constraint} & {\bf Literature} \\ \hline
Particle self-force & $k$ & $\lesssim 4.4\times 10^{48}$ & $-$\\ \hline
GUP & $\beta$ & $\lesssim 2.3\times 10^{17}$ & $\lesssim10^{43}$ \cite{Moussa} \\ \hline
Non-commutative space & $\sqrt{\theta}$ & $\lesssim 3.7\times10^6$ m& $\gtrsim 1.1\times10^{6}$ \cite{Garcia}\\ \hline
Asymptotic safety gravity & $L$ & $<7.0 \times 10^7 $ m & $-$  \\ \hline
Ho\v{r}ava-Lifshitz gravity & $\omega$ & $> 10^{-134}$ m$^{-2}$ & $\gtrsim 10^{-22}$ m$^{-2}$ \cite{Lobo} \\ \hline
\end{tabular}
\end{table}

%
\section*{Acknowledgements}

The authors  would like to thank Conselho Nacional de Desenvolvimento Cient\'{i}fico e Tecnol\'{o}gico (CNPq) for financial support.


\end{document}